\documentstyle[euscript,12pt]{article}
\catcode`@=11
\newcount\@tempcntc
\def\@citex[#1]#2{\if@filesw\immediate\write\@auxout{\string\citation{#2}}\fi
  \@tempcnta\z@\@tempcntb\m@ne\def\@citea{}\@cite{\@for\@citeb:=#2\do
    {\@ifundefined
       {b@\@citeb}{\@citeo\@tempcntb\m@ne\@citea\def\@citea{,}{\bf ?}\@warning
       {Citation `\@citeb' on page \thepage \space undefined}}%
    {\setbox\z@\hbox{\global\@tempcntc0\csname b@\@citeb\endcsname\relax}%
     \ifnum\@tempcntc=\z@ \@citeo\@tempcntb\m@ne
       \@citea\def\@citea{,}\hbox{\csname b@\@citeb\endcsname}%
     \else
      \advance\@tempcntb\@ne
      \ifnum\@tempcntb=\@tempcntc
      \else\advance\@tempcntb\m@ne\@citeo
      \@tempcnta\@tempcntc\@tempcntb\@tempcntc\fi\fi}}\@citeo}{#1}}
\def\@citeo{\ifnum\@tempcnta>\@tempcntb\else\@citea\def\@citea{,}%
  \ifnum\@tempcnta=\@tempcntb\the\@tempcnta\else
   {\advance\@tempcnta\@ne\ifnum\@tempcnta=\@tempcntb \else \def\@citea{--}\fi
    \advance\@tempcnta\m@ne\the\@tempcnta\@citea\the\@tempcntb}\fi\fi}
\catcode`@=12
\def\be{\begin{equation}}
\def\ee{\end{equation}}
\def\barr{\begin{array}}
\def\earr{\end{array}}
\def\bea{\begin{eqnarray}}
\def\eea{\end{eqnarray}}
\def\bmath{\begin{displaymath}}
\def\emath{\end{displaymath}}
\def\bq{\begin{quote}}
\def\eq{\end{quote}}
\def\oas{$\EuScript O(\alpha_s)$}

\def\g5{\gamma_5}
\def\as{\alpha_s}
\def\real{\mathop{\mbox{\rm Re}}\nolimits}

\def\chiz{\chi_{\scriptscriptstyle Z}}
\def\mz{M_{\scriptscriptstyle Z}}
\def\gz{\Gamma_{\scriptscriptstyle Z}}
\def\gf{g_{\scriptscriptstyle F}}
\def\CF{C_{\scriptscriptstyle F}}

\def\ct{\cos\theta}
\def\c2t{\cos^2\kern-2pt\theta}

\def\s2t{\sin^2\kern-2pt\theta}
\def\su{\sigma_{\scriptscriptstyle U}}
\def\sl{\sigma_{\scriptscriptstyle L}}
\def\sf{\sigma_{\scriptscriptstyle F}}
\def\stot{\sigma_{\scriptscriptstyle T}}
\def\sfb{\sigma_{\scriptscriptstyle F/0}}
\def\sfl{\sigma_{\scriptscriptstyle F/1}}

\def\Frac#1#2{\mbox{$\textstyle{#1\over#2}$}}

\def\nn{\nonumber\\}

\def\MZ{M_{\scriptscriptstyle Z}}

\def\S#1{{\EuScript S}_{#1}}

\def\SS#1{{\tilde{\EuScript S}}_{#1}}

\def\A#1#2{\mbox{$A^{#1}_{\scriptscriptstyle FB/#2}$}}
\begin{document}
\thispagestyle{empty}
\begin{flushright}
FTUV/95-67 \\[-0.2cm]
IFIC/95-70 \\[-0.2cm]
March 1996 \\[-0.2cm]
\end{flushright}
\begin{center}
{\bf{\Large Dynamical {\boldmath\oas} Effects in}}\\[.3cm]
{\bf{\Large Forward-Backward Asymmetries of}}\\[.3cm]
{\bf{\Large Heavy Quarks}}\\[1.25cm]
{\large
Michael~M.~Tung~\footnote{Feodor-Lynen Fellow}
} \\[.4cm]
Departament de F\'\i sica Te\`orica, Universitat~de Val\`encia \\
and IFIC, Centre Mixte Universitat Val\`encia --- CSIC, \\
C/ Dr.~Moliner, 50, E-46100 Burjassot (Val\`encia), Spain.
\end{center}
\vspace{5cm}
\centerline {\bf ABSTRACT}
\noindent
We examine the \oas\ forward-backward asymmetries for the production
process $e^+ e^-\to\gamma,Z\to q\bar q(g)$, tagging the outgoing
heavy-quark jet at center-of-momentum energies off the $Z$-peak.
The complicated analytic results are reduced to simple polynomial forms
that provide excellent approximations. For charm and bottom quark,
a full dynamical cancellation gives \oas\ zeros in the forward-backward
asymmetry close to the $Z$-peak. We conclude with a detailed numerical
analysis of our results.
\\[1cm]
\noindent PACS number(s): 11.38.Bx, 11.80.Fv, 14.65.-q \\
%
\noindent
The measurement of asymmetries in the production of fermion pairs at $e^+ e^-$
colliders has been proven as an indispensable tool to examine rigorously the
most important properties of the Standard Model. In general, these experimental
techniques have advanced to such an extent that theoretical predictions
beyond the Born approximation in the perturbative series of the relevant
couplings have to be taken into account in order to agree with the given
measurement precision.

Of particular interest is the forward-backward asymmetry
$A^f_{\scriptscriptstyle FB}$ which results from the vector ($V$)/axial-vector
($A$) interference terms of the intermediate $\gamma, Z$ bosons in the
production process $e^+ e^-\to\gamma,Z\to f\bar{f}$. Therefore, its measurement
allows for a direct determination of the relative strength between the $V$- and
$A$-components of the fermion coupling to the neutral current, or, equivalently,
for a precise determination of the effective electroweak mixing angle
$\s2t_{\scriptscriptstyle W}$. Experiments that measure
$A^f_{\scriptscriptstyle FB}$ for various quark flavors and leptons
are carried out by the different LEP and SLD collaborations.

This work concentrates on dynamical QCD one-loop effects in the forward-backward
asymmetries of heavy quarks. For the charm and bottom quark, high-precision
measurements on the $Z$-peak are performed by the ALEPH, DELPHI, and OPAL
collaborations at LEP~\cite{aleph,delphi,opal}. In the theoretical literature,
a first analytical treatment of massive \oas\ corrections for
$A^q_{\scriptscriptstyle FB}$ at the $Z$-peak can be found in an article
by Djouadi {\it et al.\/}~\cite{dkz}. Only recently their result (given as
an expansion in the quark mass) has been slightly corrected by Stav and
Olsen~\cite{stav}. The exact analytical formulas are lengthy and complicated,
see {\it e.g.} Ref.~\cite{tbp}.

However, in the following we shall present compact Schwinger representations
for the $\EuScript C$-odd structure function in the differential production
cross section for heavy quarks. To the best of our knowledge no such
representations have been treated in the literature before. Subsequently,
we use these results to find simple polynomial expressions for the
single-jet forward-backward asymmetry including \oas\ radiative corrections.
These approximate formulas give very accurate estimates valid over the entire
physically relevant energy spectrum, which will then allow us to reveal
interesting dynamical properties of $A^q_{\scriptscriptstyle FB}$
at QCD one-loop level. Finally, we conclude this work with a detailed
numerical analysis for charm-, bottom-, and top-quark production.

For heavy-quark production $e^+ e^-\to\gamma,Z\to q\bar{q}$, the differential
cross section is usually integrated over the azimuthal angle to yield the
following decomposition in terms of the polar angle $\theta$ of the scattered
quark
\be\label{diffrate}
{d\:\sigma\kern10pt\over d\ct} =
\Frac{3}{8}(1+\c2t)\,\su+\Frac{3}{4}\s2t\,\sl+\Frac{3}{4}\ct\,\sf.
\ee
The structure functions $\su$ and $\sl$ correspond to unpolarized
and longitudinally polarized gauge bosons, respectively, and their sum
gives the total cross section $\stot=\su+\sl$. Clearly, in Eq.~(\ref{diffrate})
the term containing $\sf$ is the only component that changes sign under
the replacement $\theta\to\pi-\theta$, and thus constitutes the odd term
under charge conjugation $\EuScript C$ in the fermionic final state.

At the Born level, $\sf$ is straightforwardly given by
\be\label{fb0}
\sfb = 8\pi\,{\alpha^2\over q^2}\,v^2 g^{V\!A},
\ee
where $q$ is the momentum transfer carried by the exchanged $\gamma$ or $Z$
boson, and $v=\sqrt{1-4m^2/q^2}$ with quark mass $m$. The factor
$g^{V\!A}$ incorporates all couplings that result from the mixed $V\!A$
interference in the intermediate state
\be\label{gva}
g^{V\!A} = -Q_q a_e a_q \real{\chiz}+2\,v_e a_e v_q a_q |\chiz|^2.
\ee
Here, the fractional charge of the quark is $Q_q$ and the relevant electroweak
coulings are $v_f=2\,T_z^f-4\,Q_f \sin^2\theta_{\scriptscriptstyle W}$ and
$a_f=2\,T_z^f$ for $f=e,q$. The $Z$-propagator displays the typical resonance
behavior for the decay of an instable massive particle so that we have
\be
\chiz(q^2)={\gf\,\mz^2\,q^2\over q^2-\mz^2+i\mz\,\gz}\quad\mbox{with}\quad
\gf={G_{\scriptscriptstyle F}\over 8\sqrt{2}\,\pi\alpha}\approx
4.299\cdot 10^{-5}\,\mbox{GeV}^{-2}.
\ee

The calculation of $\sf$ including first-order corrections in the strong
coupling involves the summation of virtual soft-gluon contributions and
hard-gluon bremsstrahlung. A closed form expression was derived in
Ref.~\cite{tbp}:
\bea\label{fb}
&& \sfl = \sfb\,\Bigg[\,
(1+\real{\tilde{A}}+\real{\tilde{C}})\,v
+{\as\over4\pi}\,\CF\,{1\over v}\,\Bigg\{\,
-(4-5\xi)\S2-\xi(1-\xi)(\SS3+\SS5) \nn[.3cm]
&&\kern1.5cm-2(4-3\xi)\S4+\xi\S6+2(\S8+\S9+3\S{10}
-\S{11})+2(1-\xi)(2-\xi)\SS{12}\Bigg\}\,
\Bigg],
\eea
where $\CF=4/3$ is the usual Casimir operator of the $SU(3)$ color group
and an additional mass parameter $\xi=1-v^2=4m^2/q^2$ has been introduced.

The coefficients $\tilde{A}$ and $\tilde{C}$ are the conventional QCD form
factors in the notation of Refs.~\cite{tbp,long} (using the Feynman gauge
and on-shell renormalization), and correspond to the gluonic
corrections of the $V$ and $A$ currents, respectively. For the axial form
factor $\tilde{C}$, it has explicitly been shown that different methods like
anticommuting $\g5$ within dimensional regularization~\cite{jlz}, dimensional
reduction~\cite{dimred}, and the 't Hooft--Veltman prescription for
$\g5$~\cite{tbp} all produce identical results.

In the following, we shall only be interested in the high-energy ($v\to0$)
and low-energy ($v\to1$) limits of the form factors $\tilde{A}$ and $\tilde{C}$
\bea
\framebox{$v\to1$} &
\real{\tilde{A}}\sim\real{\tilde{C}}\sim &
{\as\over4\pi}\CF\Bigg[\,\ln^2\xi-(1+4\ln2)\ln\xi+2(1+2\ln2)\ln2 \nn[.3cm]
&&\kern1.4cm+\Frac{4}{3}\pi^2-4+\EuScript O(\xi)\,\Bigg], \label{ac1} \\[.3cm]
\framebox{$v\to0$} &
\real{\tilde{A}}\sim &
{\as\over4\pi}\CF\Bigg[\,{\pi^2\over v}-8+\pi^2 v+\Frac{2}{9}(1+24\ln2)v^2+
\EuScript O(v^4)\,\Bigg], \label{a0} \\[.3cm]
\framebox{$v\to0$} &
\real{\tilde{C}}\sim &
{\as\over4\pi}\CF\Bigg[\,{\pi^2\over v}-4+\pi^2 v-\Frac{2}{9}(11-24\ln2)v^2+
\EuScript O(v^4)\,\Bigg]. \label{c0}
\eea
Eq.~(\ref{ac1}) clearly states that for nearly massless quarks or sufficiently
high center-of-momentum energies, $E_{cms}=\sqrt{q^2}$, the virtual-gluon
corrections become insensitive to the parity property of the relevant
vector-boson vertex. However, in the asymptotic energy range near threshold,
$\tilde{A}$ and $\tilde{B}$ differ by a finite contribution reflecting the
distinct nature of the underlying symmetries. Following the reasoning by
Schwinger~\cite{schw}, both $1/v$-poles in Eqs.~(\ref{a0}) and (\ref{c0})
correspond to the strong attraction between the color charges (quarks) in
the non-relativistic limit with a relative velocity $2v\ll1$.

In Eq.~(\ref{ac1}) the collinear IR divergences emerge as logarithmic
singularities for $\xi\to0$, and eventually cancel when the hard-gluon
parts are added. (Note that the trivial soft IR divergences have already
been neglected as indicated by the wiggle, {\it viz.} Ref.~\cite{tbp}.)
The full analytic solutions of the $q\bar{q}g$ phase-space integrals
$\S{i}$, $i=2,\ldots,12$, have been calculated and classified in
Ref.~\cite{long}. Much simpler results which approximate the exact
solutions in the important high- and low- energy domains were presented in
           Ref.~\cite{mmt}~\footnote{
           Note the typographical error in Table~I of
           Ref.~\cite{mmt}. The correct limiting
           behavior of the `spin-dependent' integral
           $\SS5$ close to threshold ($v\to0$) is
           $\SS5\sim4\big(\,2\ln v-1\,\big)+\EuScript O(v)$.}.

It is now straightforward to take the limits in the total expression
Eq.~(\ref{fb}). For $v\to1$, we obtain
\be\label{sfto1}
\sfl\sim\sfb\Big[\,1+\as\,\EuScript O\big(1-v^2\big)\,\Big],
\ee
{\it i.e.} the QCD corrections for $\sf$ are genuine quark-mass effects
and can safely be neglected in the high-energy region. This prediction
was already made in Ref.~\cite{jlz}. On the other hand, the asymptotic
behavior close to threshold has never before been considered in the
literature. Our result for $v\to0$ is
\be\label{sfto0}
\sfl\sim\sfb\left[\,1+{\as\over2\pi}\,\CF\left\{
{\pi^2\over v}-6+\pi^2v+\EuScript O(v^2)\right\}\,\right].
\ee
Although non-perturbative resonance effects at threshold supersede
predictions made by perturbation theory, Eq.~(\ref{sfto0}) gives
essential information on $\sf$ sufficiently above the production resonance.

Combining Eqs.~(\ref{sfto1}) and (\ref{sfto0}) yields the following
Schwinger representation for $\sf$
\be\label{phirep}
{\sfl\over\sfb} = 1+\CF\,\as\,{\pi\over2}\left(\,{1\over v}-\varphi(v)\,\right),
\ee
with the mass-zero condition $\varphi(1)\equiv1$. This mass-zero condition
is an absolute requirement for any $\varphi$-representation, whereas the
exact threshold value $\varphi(0)=6/\pi^2$ is of lesser importance due to the
$1/v$-pole dominance in Eq.~(\ref{phirep}). In general, $\varphi(v)$ is a
function of $0\le v\le1$ and contains apart from the universal term in
Eq.~(\ref{phirep}) (describing the color interaction of the quarks close
to threshold) all the non-trivial energy dependence of $\sfl$. Suitable
$\varphi$-representations that provide excellent approximations to the
exact solutions are simple polynomials of degree $m\ge2$:
\be
\varphi_m(v) =
{\sum\limits_{i=0}^m\,a_i\,v^i\over\sum\limits_{i=0}^m\,a_i}.
\ee
Tab.~1 displays the coefficients $a_i$ for the lowest-order representations
$\varphi_2$, $\varphi_3$, and $\varphi_4$. In Fig.~1, we have plotted these
polynomial representations together with the exact result. Already the
second-order form $\varphi_2$ gives a very accurate interpolation so that
the corresponding Schwinger formula Eq.~(\ref{phirep}) provides a very compact
expression for easy implemention of the \oas\ corrections to $\sf$.

A straightforward procedure to include these massive \oas\ effects already
in the Born approximation consists in the introduction of effective couplings.
From Eqs.(\ref{gva}) and (\ref{phirep}) we obtain directly the prescription
\be\label{gvaeff}
g^{V\!A}\:\to\:
\tilde{g}^{V\!A} = g^{V\!A}\,\left[\,
1+\CF\,\as\,{\pi\over2}\left(\,{1\over v}-\varphi_m(v)\,\right)
\,\right],
\ee
where $\varphi_m(v)$ are the appropriate polynomials of Tab.~1.

Similarly, one finds replacements rules for the two remaining couplings,
$g^{VV}$ and $g^{AA}$, that multiply with the $\EuScript C$-even components
of the differential rate. They contribute through the $VV$ and $AA$
parity-parity combinations of the intermediate $\gamma,Z$ states to the
to the total rate
\be\label{stot}
\stot = {4\pi\alpha^2\over q^2}\,\left[\,
\Frac{1}{2}v\,(3-v^2)\,g^{VV}+v^3g^{AA}\,\right].
\ee
Using the third-order Schwinger representions given in Ref.~\cite{mmt}, we
find the following explicit \oas\ expressions
\bea\label{gvveff}
\tilde{g}^{VV} &=&
\Big[\,Q_q^2-2\,Q_q v_e v_q \real{\chiz}+(v_e^2+a_e^2)\,v_q^2\,|\chiz|^2\,\Big]
\ \times\nn[.3cm]
&& \left[\,1+\CF\,\as\left\{{\pi\over2v}-
\left({\pi\over2}-{3\over4\pi}\right)
{95-82v+173v^2-85v^3\over101}  
\right\}\,\right], 
\eea
and
\bea\label{gaaeff}
\tilde{g}^{AA} &=&
(v_e^2+a_e^2)\,a_q^2\,|\chiz|^2
\ \times\nn[.3cm]
&& \left[\,1+\CF\,\as\left\{{\pi\over2v}-
\left({\pi\over2}-{3\over4\pi}\right)
{43-30v+15v^2+71v^3\over99}  
\right\}\,\right].
\eea

Now all ingredients are available to treat the forward-backward asymmetries
of heavy quarks at QCD one-loop level. The forward backward-asymmetry
$A^f_{\scriptscriptstyle FB}$ measures the fermion events in the forward
and backward hemispheres and is therefore defined as
\be
A^f_{\scriptscriptstyle FB} \ =\
{\displaystyle
 \int\limits_0^1 d\!\ct\ {d\:\sigma\kern10pt\over d\ct} -
 \int\limits_{-1}^0 d\!\ct\ {d\:\sigma\kern10pt\over d\ct}\over
 \displaystyle
 \int\limits_0^1 d\!\ct\ {d\:\sigma\kern10pt\over d\ct} +
 \int\limits_{-1}^0 d\!\ct\ {d\:\sigma\kern10pt\over d\ct}}\ ,
\ee
which immediately gives with Eq.~(\ref{diffrate})
\be\label{affb}
A^f_{\scriptscriptstyle FB} \ =\
{3\over4}\,{\sf\over\su+\sl}.
\ee
For heavy-quark production, the \oas\ corrections for
$A^q_{\scriptscriptstyle FB}$ take a particularly simple form when
expressed in terms of the effective couplings $\tilde{g}^{ij}$, with
$i,j=V,A$. Using Eqs.~(\ref{fb0}) and (\ref{stot}) we obtain
\be\label{afb1}
\A{q}{1} =
{3v\,\tilde{g}^{V\!A}\over(3-v^2)\,\tilde{g}^{VV}+2v^2\tilde{g}^{AA}}.
\ee
Close to the $Z$-peak, the $|\chiz|^2$ propagator terms in the couplings
dominate and the corresponding Born expression $\A{q}{0}$ reduces to the
approximate formula given in Ref.~\cite{bodo}.

Fig.~2 shows the energy dependence of $A^q_{\scriptscriptstyle FB}$ for
charm, bottom, and top quark in the Born approximation (solid line) and
including the \oas\ final state corrections (dashed line). For the running
of the strong coupling we have used the central value $\as^{(5)}(\MZ)=0.118$
with five active flavors on the $Z$-peak. Crossing the top-quark threshold
the appropriate matching condition has to be imposed. For charm and bottom
quark, the fixed-point masses $m_c(m_c)=1.3$~GeV and $m_b(m_b)=4.33$~GeV
correspond to $m_c(\MZ)=0.78$~GeV and $m_b(\MZ)=3.20$~GeV, respectively.
In the top-quark case, we have chosen $M_t=180$~GeV as pole mass, which
gives the fixed-point mass $m_t(m_t)=172.1$~GeV. In each individual plot
of Fig.~2, the dashed-dotted line refers to the right-hand scale measuring
the absolute difference between Born and \oas\ predictions, {\it i.e.\/}
$\Delta A^q_{\scriptscriptstyle FB} = \A{q}{1}-\A{q}{0}$. For certain
center-of-momentum energies a subtle cancellation of the \oas\ expressions
in the numerator and denominator of Eq.~(\ref{affb}) takes place and
gives dynamical zeros for $\Delta A^q_{\scriptscriptstyle FB}$.

Fig.~2(a) highlights the two specific energy values $E_{cms}=6.58$~GeV and
$E_{cms}=89.362$~GeV at which the massive one-loop QCD corrections vanish.
At the first energy point, we find
$\A{c}{0}(6.58~\mbox{GeV})=\A{c}{1}(6.58~\mbox{GeV})=0.0038$.
Of particular interest is the energy domain close to 89.362~GeV,
where both lowest-order as well as \oas\ one-loop contributions to the
forward-backward asymmetry vanish. These cancellations originate from the
dynamical interplay of the involved electroweak couplings at Born level.
Note that the \oas\ result factorizes with the Born term so that the isolated
one-loop zero occurs only for the lower energy value $E_{cms}=6.58$~GeV.

As illustrated in Fig.~2(b), a similar situation arises for the bottom quark.
Here, our theoretical predictions for the dynamical zeros are
$\A{b}{0}(23.32~\mbox{GeV})=-0.1015$ with
$\Delta A^b_{\scriptscriptstyle FB}(23.32~\mbox{GeV})=0$,
and $\A{b}{0}(85.235~\mbox{GeV})=\A{b}{1}(85.235~\mbox{GeV})=0$.

An interesting observation is that for charm and bottom quark the lower
energy values yielding dynamical zeros correspond both to approximately
the same mass scale $v=0.9512$. Note that on the $Z$-peak the predictions
of Figs.~2(a) and (b) agree very well with the recent estimates of
Ref.~\cite{bodo}.

Due to the high production threshold of the top quark, only one first-order
$\as$ zero is located at 848.80~GeV in Fig.~2(c). The corresponding
forward-backward asymmetry at Born level is $\A{t}{0}=0.5557$.

To isolate the energy dependence of the non-trivial one-loop contents
in $\A{q}{1}$, we introduce the function $\Phi^q(v)$ in the following
manner
\be
\A{q}{1} = \A{q}{0}\,\left[\,1+{\as\over\pi}\,\Phi^q(v)\,\right].
\ee
Note that $\Phi^q(v)$ contains additional couplings $g^{VV}$ and $g^{AA}$,
which depend solely on $v$ once the quark type $q$ is fixed and the
corresponding running mass $m_q(q^2)$ is implemented.

In the following, we shall use the notation $\Phi^q$ for the exact \oas\
result derived from the analytic expressions for $\sf$ and $\stot$ as given in
Eq.~(\ref{fb}) and in Ref.~\cite{mmt}, respectively. Any additional subscript
will indicate approximate representations different from the exact formula
$\Phi^q$.

The exact result $\Phi^q$ involves all the analytic integral solutions
listed in Ref.~\cite{tbp} yielding complicated and lengthy expressions.
However, using simple polynomial interpolations for the couplings
$\tilde g^{ij}$, we can obtain compact and very accurate approximations
that allow for an easy implementation of the forward-backward asymmetry.
For example, the substitution of the $\EuScript O(v^3)$ expressions
Eqs.~(\ref{gvaeff}), (\ref{gvveff}), and (\ref{gaaeff})) into
Eq.~(\ref{afb1}) gives such a useful representation, which we denote
as $\Phi^q_3$.

Further simplification is accomplished by expanding numerator and denominator
of Eq.~(\ref{afb1}) up to first oder in $\as$. After inserting the numerical
values for the relevant electroweak couplings and the running of the quark
masses, we find the following lowest-order ($L$) interpolations
\bea
\Phi^{b,c}_{\scriptscriptstyle L}(v) &=&
-191.175 + 413.338\,v - 223.098\,v^2, \\
\Phi^t_{\scriptscriptstyle L}(v) &=&
-0.427 + 5.187\,v - 5.070\,v^2.
\eea
Note that in this representation $\Phi_{\scriptscriptstyle L}$ is practically
insensitive to the distinct couplings and masses of bottom and charm quark,
{\it i.e.} $\Phi^{b,c}_{\scriptscriptstyle L}(1)\approx-1$.
Only the top quark requires a separate parametrization due to its exceptionally
high mass.

To illustrate the high accuracy of these compact formulas, we have plotted
in Figs.~3 the representations $\Phi_3$ and $\Phi_{\scriptscriptstyle L}$,
and the exact \oas\ result $\Phi$ for charm, bottom, and top quark. Furthermore,
we have also included the following approximation obtained from an expansion of
the form factors to second order in $\sqrt{\xi}$ (first order in $m$) as given
in Ref.~\cite{stav}
\be
\Phi^q_{\scriptscriptstyle Z} =
-1+\Frac{8}{3}\sqrt{\xi}+\Frac{1}{3}\xi\left[7+\Frac{1}{6}\pi^2-
2\ln\left(\Frac{1}{2}\sqrt{\xi}\right)+\ln^2\left(\Frac{1}{2}\sqrt{\xi}\right)
\right]
-3\,\xi\,{v_q^2-2a_q^2 \ln\left(\Frac{1}{2}\sqrt{\xi}\right)\over v_q^2+a_q^2}.
\ee
Note that $\Phi_{\scriptscriptstyle Z}$ contains a part which explicitly
depends on the electroweak couplings. This formula has primarily been used
for approximate estimates on the $Z$-peak.

Comparing Figs.~3(a) and (b), we recognize that in the charm- and bottom-quark
case for $v>0.8$ all specific properties as flavor and mass effects fully
factorize with the Born contribution $\A{c,b}{0}$, and $\Phi$ exhibits the
same universal functional dependence. In the region $v>0.9$,
which corresponds to center-of-momentum energies above 5 and 20 GeV for charm
and bottom quark, respectively, $\Phi_{\scriptscriptstyle L}$ gives the best
representation. Closer to threshold, $\Phi_3$ provides more accurate numerical
estimates.

Fig.~3(c) demonstrates that for the top quark $\Phi_{\scriptscriptstyle L}$
and $\Phi_{\scriptscriptstyle 3}$ yield far better results than
$\Phi_{\scriptscriptstyle Z}$. Higher-order mass terms in
$\Phi_{\scriptscriptstyle Z}$ should be taken into account to reach
a comparable precision.

To complete this discussion, we present in Tabs.~2--4 the explicit numerical
values of Born and \oas\ contributions to the forward-backward asymmetry
for charm, bottom, and top quark with the same choice for $\as(\MZ)$ and
the fixed-point quark masses $m_q(m_q)$ as in Fig.~2. In the fourth column
the impact of the \oas\ contributions on the Born result is given by
$(\A{q}{1}-\A{q}{0})/\A{q}{0}$ in percent. Again, $\A{q}{3}$,
$\A{q}{L}$, and $\A{q}{Z}$ denote the estimates stemming from $\Phi^q_3$,
$\Phi^q_{\scriptscriptstyle L}$, and $\Phi^q_{\scriptscriptstyle L}$,
respectively. For these approximate results we also provide in the adjacent
columns the relative error to $\A{q}{1}$ in percent. By comparing these
relative errors with the corresponding percentages capturing the exact
\oas\ modifications, one obtains a reliable measure of the quality of
the approximations.

In this work, we have investigated the QCD one-loop corrections to the
single-jet forward-backward asymmetries for massive quarks off the
$Z$-resonance, which modify the Born predictions by approximately 3\%.
After finding compact Schwinger representations for the $\EuScript C$-odd
contribution in the angular distribution of the produced quark, we derived
simple polynomial expressions that provide excellent approximations for
the \oas\ forward-backward asymmetry.

Due to a dynamical cancellation of $\EuScript C$-even and $\EuScript C$-odd
components in the single-jet asymmetry, \oas\ zeros occur close to the $Z$-peak
at 89.362~GeV for charm production and at 85.235~GeV for bottom production.
A thorough experimental investigation of this effect should reveal interesting
new insights in the physics at higher-order QCD level and might even allow
for a detection of signals beyond the Standard Model.
\vskip1cm\noindent
{\bf Acknowledgements.} It is a pleasure to thank J.~Bernab\'eu and
J.~Pe\~narrocha for a critical reading of the manuscript, and
K.~Kleinknecht for a helpful conversation. I further acknowledge
the support given by the Alexander-von-Humboldt Foundation and
CICYT under grant AEN93-0234.
\newpage
\newpage
\centerline{\bf\Large Figure Captions }
\vspace{1cm}
\newcounter{fig}
\begin{list}{\bf\rm Fig.\ \arabic{fig}: }{\usecounter{fig}
\labelwidth1.6cm \leftmargin2cm \labelsep0.4cm \itemsep0ex plus0.2ex }
\item Schwinger representations for the QCD one-loop corrections
      to $\sf$ in the differential rate of the production process
      $e^+ e^-\to\gamma,Z\to q\bar q(g)$. The exact result is compared
      to the polynomial representations of degree $m=2,3$, and 4.
\item The single-jet forward-backward asymmetry
      $A^q_{\scriptscriptstyle FB}$ in the Born approximation and
      at \oas\ as a function of the center-of-momentum energy for
      (a) charm, (b) bottom, and (c) top production. The dashed-dotted line
      refers to the right ordinate where $\Delta A^q_{\scriptscriptstyle FB}$
      denotes the difference between Born and \oas\ results.
\item Non-trivial one-loop contents of $\A{q}{1}$ as a function of
      $v=\sqrt{1-4m_q^2/q^2}$ for (a) charm, (b) bottom, and (c) top
      quark. The exact result $\Phi^q$ is compared to the representations
      $\Phi_3$, $\Phi_{\scriptscriptstyle L}$, and
      $\Phi_{\scriptscriptstyle Z}$.
\end{list}
\newpage
\centerline{\bf\Large Table Captions }
\vspace{1cm}
\newcounter{tab}
\begin{list}{\bf\rm Tab.\ \arabic{tab}: }{\usecounter{tab}
\labelwidth1.6cm \leftmargin2cm \labelsep0.4cm \itemsep0ex plus0.2ex }
\item Coefficients for the lowest-order Schwinger representations of $\sf$.
\item Forward-backward asymmetry for $e^+ e^-\to\gamma,Z\to c\bar c$ at
      Born level and up to \oas\ compared to estimates given by $\Phi_3$,
      $\Phi_{\scriptscriptstyle L}$, and $\Phi_{\scriptscriptstyle Z}$
      ($\as(\MZ)=0.118$ and $m_c(m_c)=1.3$~GeV).
\item Forward-backward asymmetry for $e^+ e^-\to\gamma,Z\to b\bar b$ at
      Born level and up to \oas\ compared to estimates given by $\Phi_3$,
      $\Phi_{\scriptscriptstyle L}$, and $\Phi_{\scriptscriptstyle Z}$
      ($m_b(m_b)=4.33$~GeV).
\item Forward-backward asymmetry for $e^+ e^-\to\gamma,Z\to t\bar t$ at
      Born level and up to \oas\ compared to estimates given by $\Phi_3$,
      $\Phi_{\scriptscriptstyle L}$, and $\Phi_{\scriptscriptstyle Z}$
      ($M_t=180$~GeV).
\end{list}

\begin{thebibliography}{99}
\bibitem{aleph}  The ALEPH Collaboration, D.~Busculic {\it et al.\/},
                 Phys.\ Lett.\ B {\bf 352}, 479 (1995),
                 and Z.\ Phys.\ C {\bf 62}, 179 (1994).
\bibitem{delphi} The DELPHI Collaboration, P.~Abreu {\it et al.\/},
                 Z.\ Phys.\ C {\bf 66}, 341 (1995).
\bibitem{opal}   The OPAL Collaboration, G.~Alexander {\it et al.\/},
                 {\it Measurement of the Heavy Quark Forward-Backward
                 Asymmetries and Average B Mixing Using Leptons in
                 Multi-Hadronic Events}, CERN Report No.\
                 CERN-PPE/95-179, December 1995 (unpublished), and
                 R.~Akers {\it et al.\/}, Z.\ Phys.\ C {\bf 68}, 203 (1995).
\bibitem{dkz}    A.~Djouadi, J.H.~K\"uhn, and P.M.~Zerwas, Z.\ Phys.\ C
                 {\bf 46}, 411 (1990).
\bibitem{stav}   J.B.~Stav and H.A.~Olsen, Phys.\ Rev.\ D {\bf 52}, 1359 (1995).
\bibitem{tbp}    M.M.~Tung, J.~Bernab\'eu, and J.~Pe\~narrocha, {\it Analytic
                 \oas\ Results for Bottom and Top Quark Production in $e^+e^-$
                 Collisions}, preprint no.\ FTUV/95-38 and IFIC/95-40,
                 November 1995 (hep-ph/9601277).
\bibitem{long}   J.G.~K\"orner, A.~Pilaftsis and M.M.~Tung, Z.\ Phys.\ C
                 {\bf 63}, 575 (1994) (hep-ph/9311332); M.M.~Tung, Ph.D.\
                 thesis, University of Mainz, 1993.
\bibitem{jlz}    J.~Jer\'sak, E.~Laermann and P.M.~Zerwas, Phys.\ Lett.\ B
                 {\bf 98}, 363 (1981); Phys.\ Rev.\ D {\bf 25} 1218 (1982).
\bibitem{dimred} J.G.~K\"orner and M.M.~Tung, Z.\ Phys.\ C {\bf 64}, 255 (1994).
\bibitem{schw}   J.~Schwinger, {\it Particles, Sources and Fields}
                 (Addison-Wesley, Redwood City, 1988), Vol.~III, pp.~99.
\bibitem{mmt} M.M.~Tung, Phys.\ Rev.\ D {\bf 52}, 1353 (1995) (hep-ph/9403322).
\bibitem{bodo}   A.~Djouadi, B.~Lampe, and P.M.~Zerwas, Z.\ Phys.\ C
                 {\bf 67}, 123 (1995) (hep-ph/9411386).
%
%
\end{thebibliography}
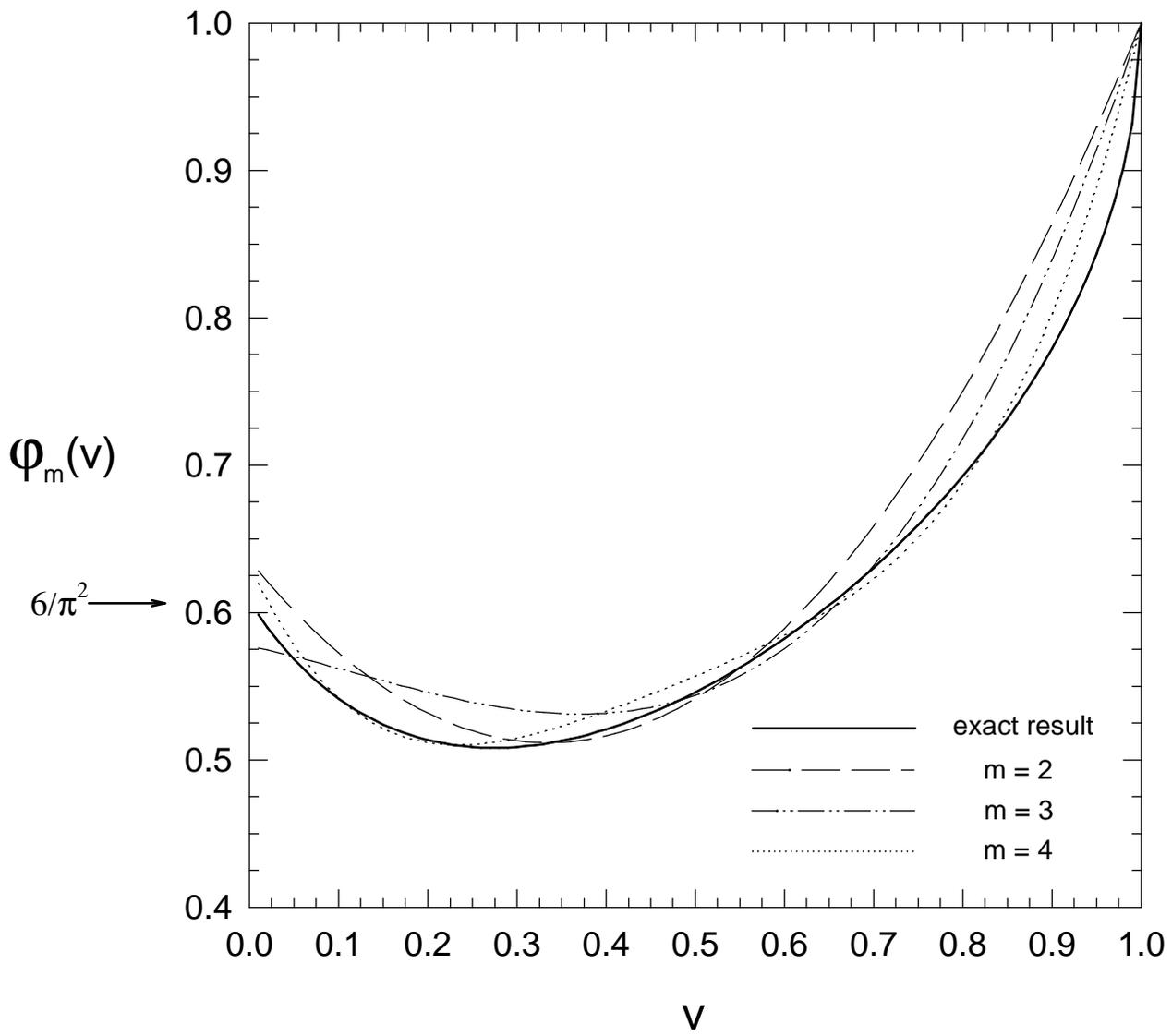
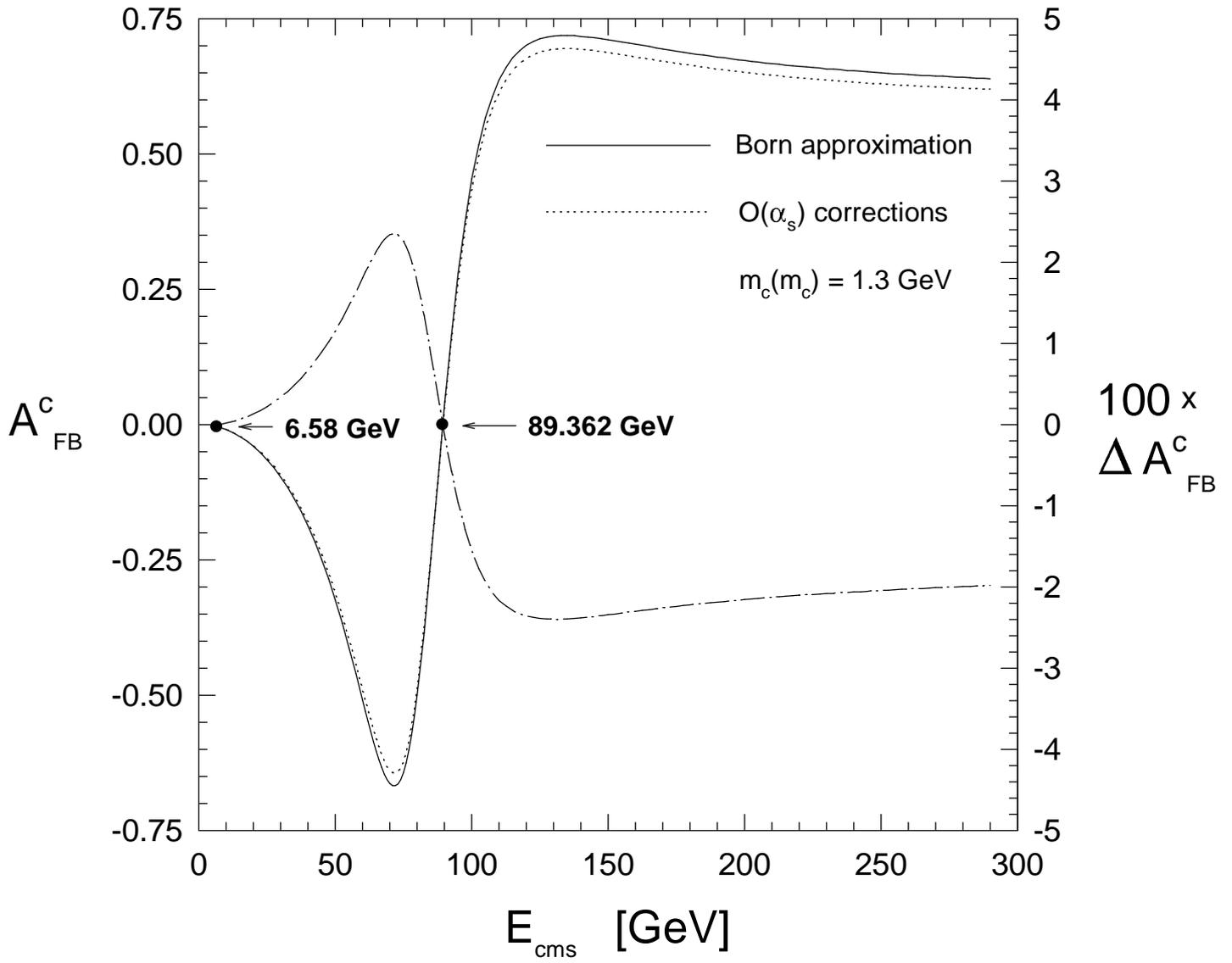
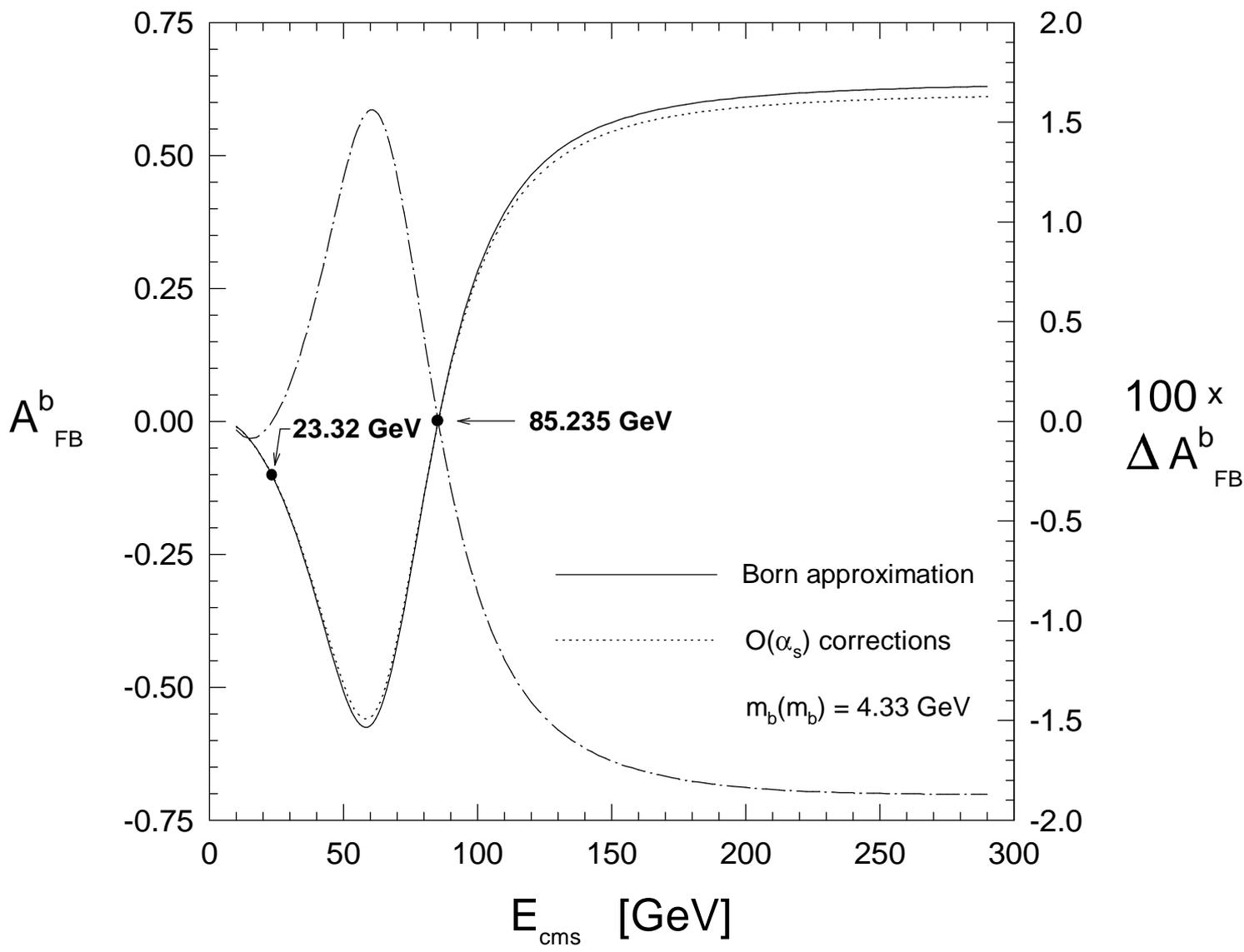
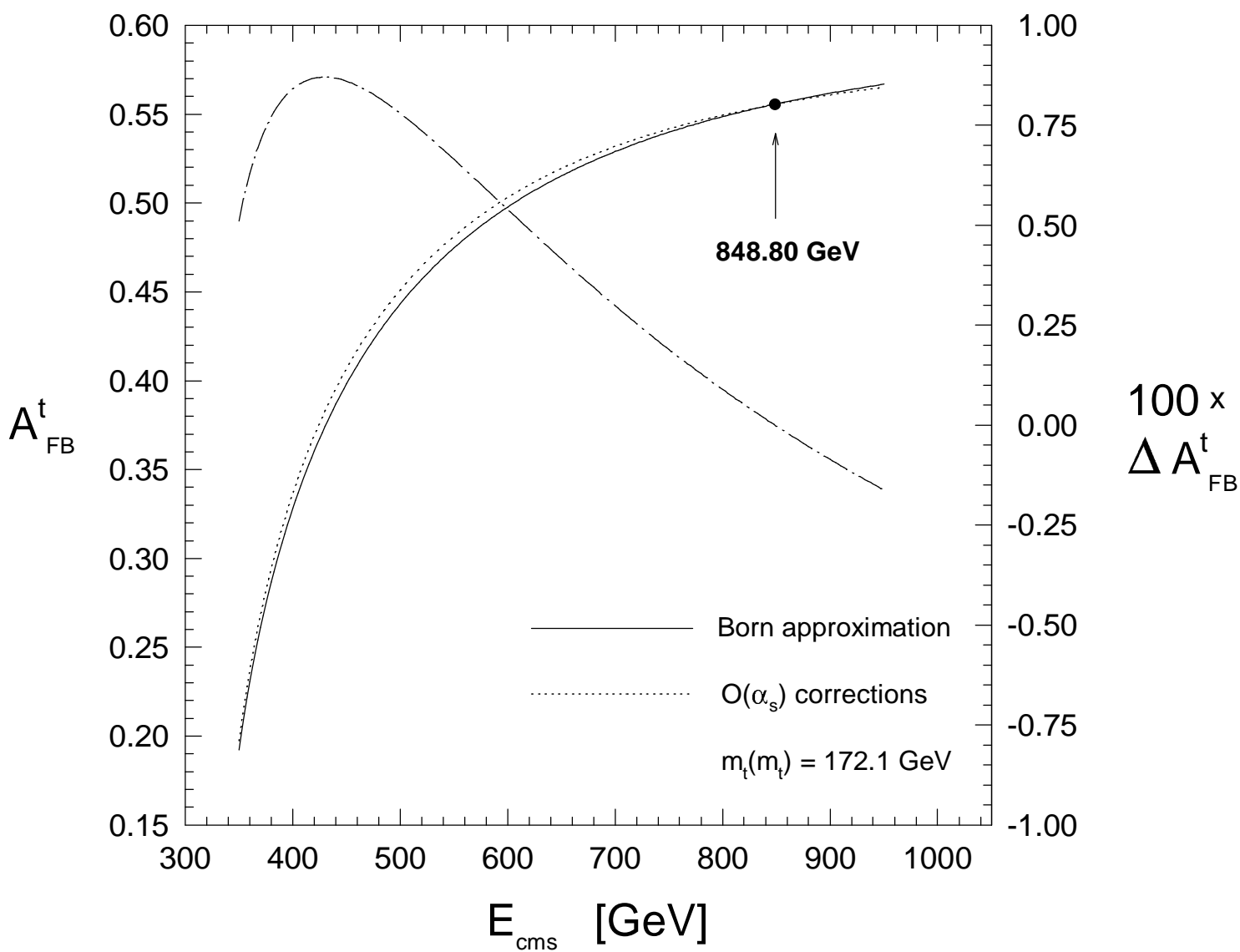
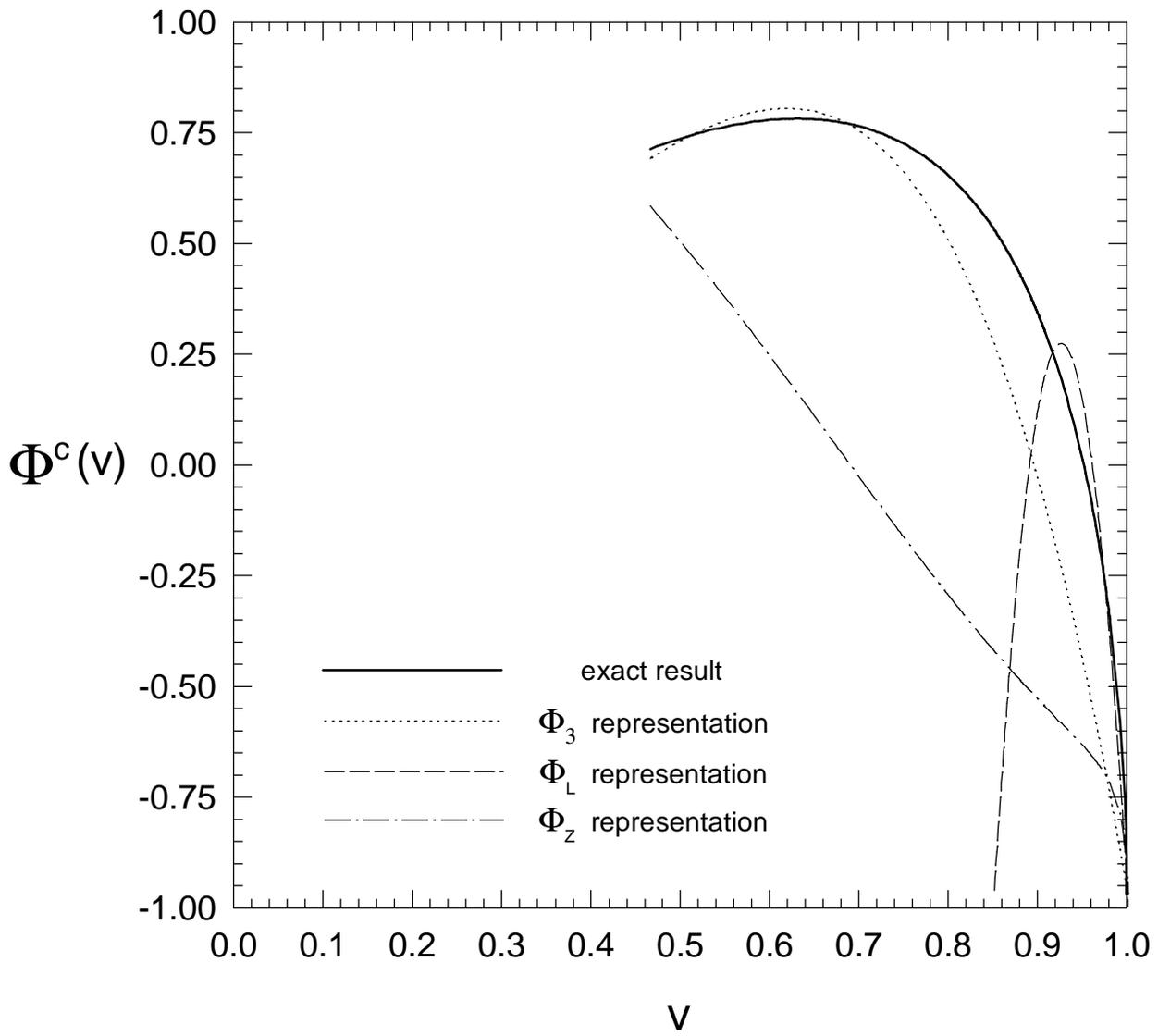
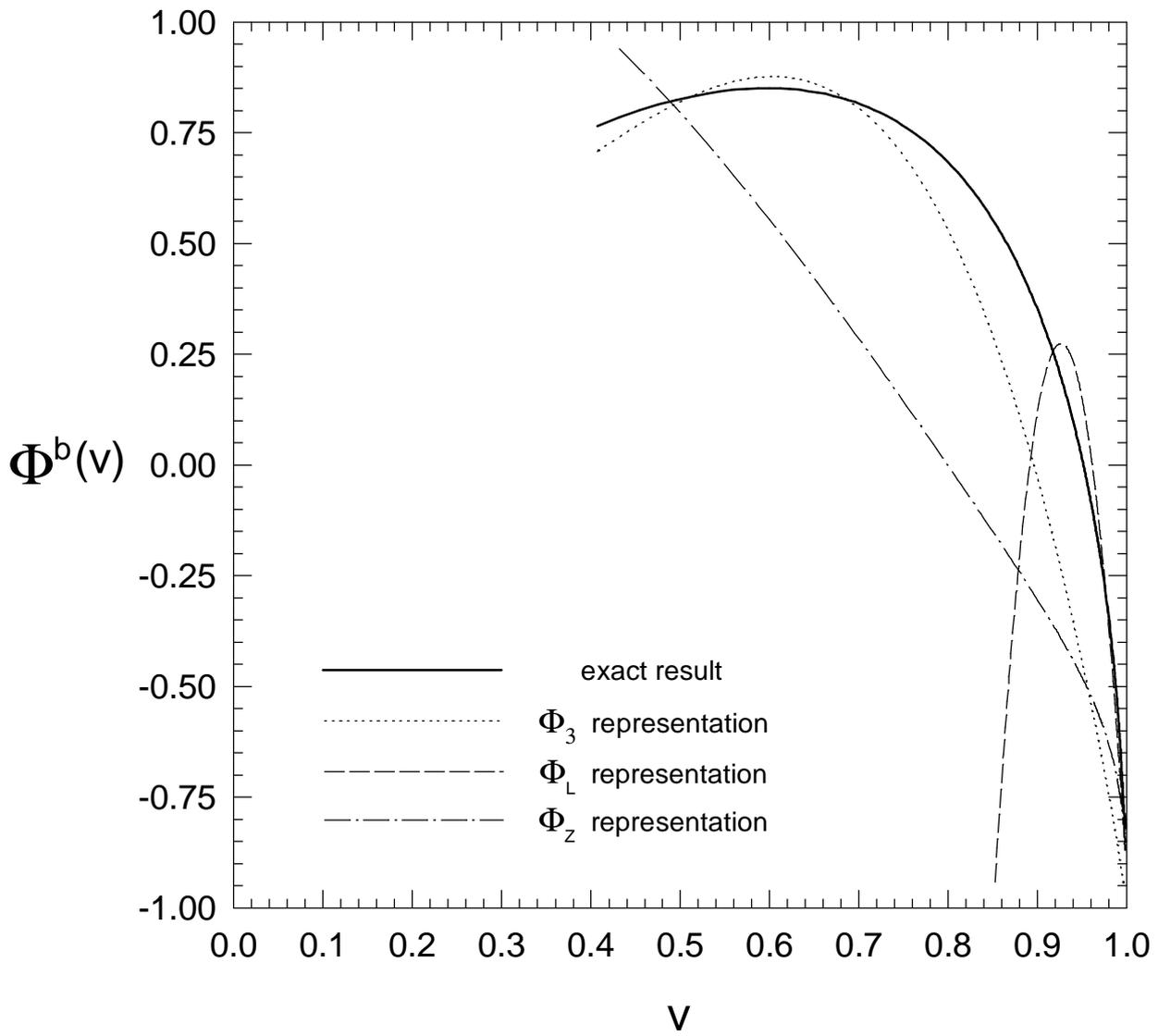
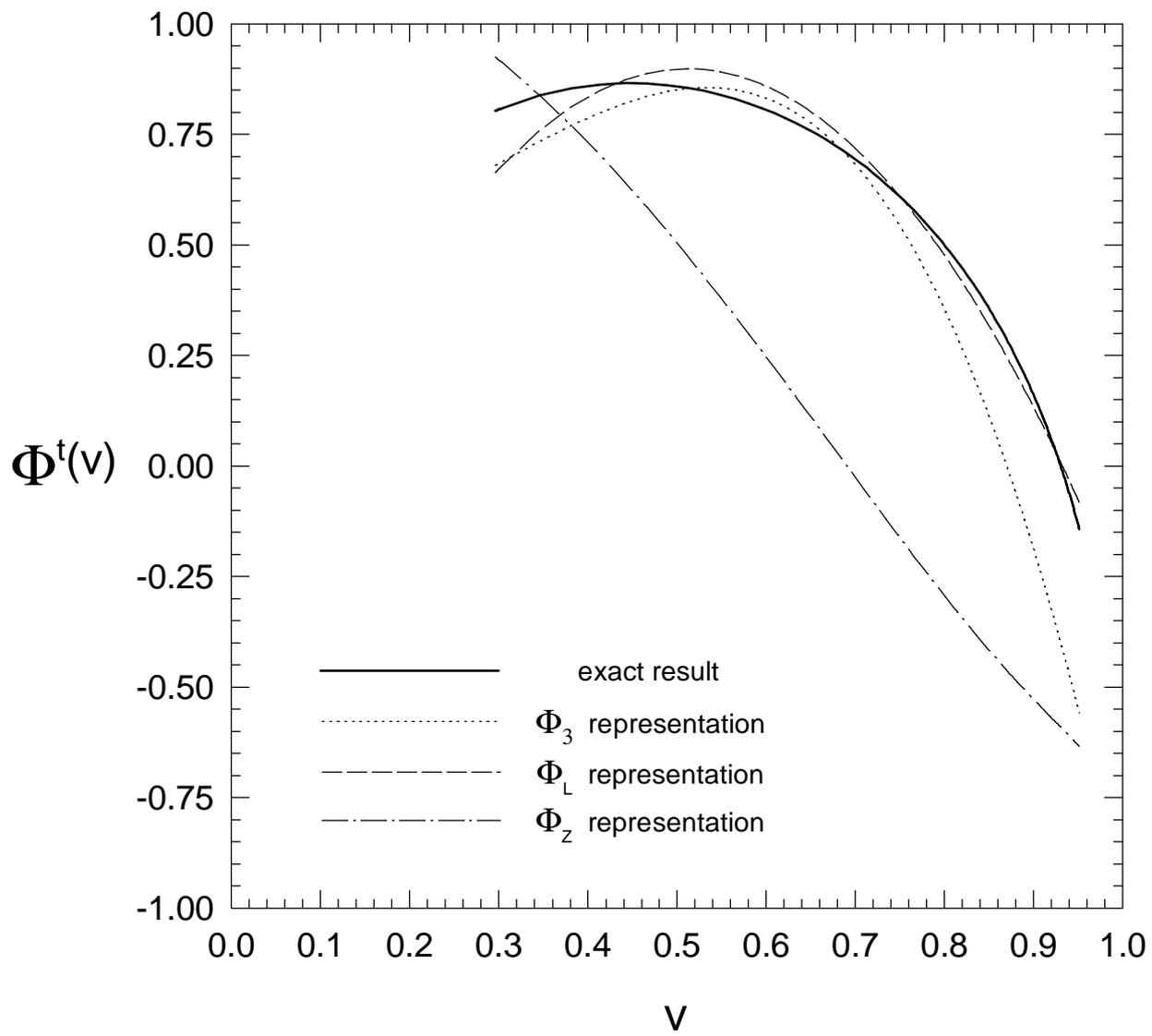
\end{document}